# Effects of proton irradiation on 60 GHz CMOS transceiver chip for multi-Gbps communication in high-energy physics experiments

Imran Aziz[1,2] ✉, Dragos Dancila[1], Sebastian Dittmeier[3], Alexandre Siligaris[4], Cedric Dehos[4], Patrik Martin De Lurgio[5], Zelimir Djurcic[5], Gary Drake[5], Jose Luis Gonzalez Jimenez[4], Leif Gustaffson[1], Don-Won Kim[6], Elizabeth Locci[7], Ulrich Pfeiffer[8], Pedro Rodriquez Vazquez[8], Dieter Röhrich[9], Andre Schöning[3], Hans Kristian Soltveit[3], Kjetil Ullaland[9], Pierre Vincent[4], Shiming Yang[9], Richard Brenner[1]

[1]*Uppsala University, Sweden*
[2]*Mirpur University of Science & Technology (MUST), Pakistan*
[3]*Heidelberg University, Germany*
[4]*CEA-Leti, France*
[5]*Argonne Laboratory, USA*
[6]*Gangneung-Wonju University, Republic of Korea*
[7]*CEA/DSM/IRFU/DphP & Paris-Saclay University France*
[8]*Wuppertal University, Germany*
[9]*Bergen University, Norway*
✉ *E-mail: imran.aziz@angstrom.uu.se*

**Abstract:** This article presents the experimental results of 17 MeV proton irradiation on a 60 GHz low power, half-duplex transceiver (TRX) chip implemented in 65 nm CMOS technology. It supports short range point-to-point data rate up to 6 Gbps by employing on-off keying (OOK). To investigate the irradiation hardness for high-energy physics (HEP) applications, two TRX chips were irradiated with total ionising doses (TID) of 74 and 42 kGy and fluence of $1.4 \times 10^{14} N_{eq}/cm^2$ and $0.8 \times 10^{14} N_{eq}/cm^2$ for RX and TX modes, respectively. The chips were characterised by pre- and post-irradiation analogue voltage measurements on different circuit blocks as well as through the analysis of wireless transmission parameters like bit error rate (BER), eye diagram, jitter etc. Post-irradiation measurements have shown certain reduction in performance but both TRX chips have been found operational through over the air measurements at 5 *Gbps*. Moreover, very small shift in the carrier frequency was observed after the irradiation.

## 1 Introduction

The upgrade of accelerators and experiments at the Large Hadron Collider (LHC) for high luminosity will result in multiple times higher event rates, which demands high data rate readout systems [1]. For instance, the readout data requirement at the upgraded ATLAS silicon micro-strip tracker will be from 50 to 100 Tbps [2]. The transfer rate for highly granular tracking detectors is limited by the available bandwidth in current readout links [3]. Current readout systems typically use wired optical links which are immune to cross-talk but require large connector size and are sensitive to mechanical damage.

Wireless data transmission using millimetre-wave technology has not yet been used in trackers for particle physics experiments. The wireless allowing data and power transmission (WADAPT) collaboration [4] was formed to study the feasibility of wireless technologies in high-energy physics (HEP) experiments. The objective is to provide a common platform for research and development in order to optimise the effectiveness and cost, with the aim of designing and testing wireless demonstrators for large instrumentation systems. A strong motivation for using wireless data transmission is the absence of wires and connectors which will be an advantage in areas with low material budget. The availability of 9 GHz license-free bandwidth at 60 GHz (57–66 GHz) provides the opportunity to achieve multi-Gbps data rates for a single link. In addition, wireless transmission allows radial readout instead of current axial links, which opens for topological readout of tracking detectors. This can also reduce on-detector data volume using inter-layer communication, which will drastically reduce the readout time and latency [5].

Technology used in HEP experiments, must withstand high radiation levels. The current detectors at LHC have accumulated radiation levels of $\sim 2 \times 10^{15} N_{eq}/cm^2$ and total ionising dose (TID) of $\sim 300$ kGy. These values are expected to increase tenfold for the high luminosity upgrade of LHC planned to be operational in 2026 [6]. With this target application in mind, we have performed an initial investigation into the radiation hardness of a low power, fully integrated 60 GHz wireless transceiver (TRX) chip. The TRX is fabricated in 65 nm CMOS technology targeting mobile devices requiring multi-Gbps data transfer. The chips were not designed for use in the radiation environment; hence, there were no guarantees for any functionality even after low doses. The irradiation was, therefore, done with the chips in operation to see in real time if the chips fail. The complexity of the setup limited the dose in this study, however, relevant results were obtained up to 74 kGy.

## 2 60 GHz CMOS transceiver

The TRX chip under investigation has been manufactured by STMicroelectronics in 65 nm CMOS technology. This V-band short-range chip can support point-to-point data rates from 10 Mbps to 6 Gbps by employing an on-off keying (OOK) modulation scheme. Moreover, it operates in half-duplex mode with power consumption of 40 mW in transmit mode, 25 mW in receive mode, and 10 μW in idle mode. The chip is packaged as a 25-ball very thin profile, fine-pitch ball grid array (VFBGA) with the overall dimensions of 2.2 mm x 2.2 mm x 1.0 mm. The low energy consumption as well as low profile of the chip make it a good candidate for the targeted applications. A modulation block diagram of the chip is shown in Fig. 1*a*. The carrier is generated in the transmitter subsystem by an on-chip voltage-controlled oscillator (VCO), which provides a differential sine-wave signal around 60.5 GHz. This signal is amplified to provide a large sine-wave to the modulator, while the amplification is controlled through feedback calibrated path. The amplified signal, along with the differential data from a line receiver, are fed to the OOK modulator. The output of the modulator is then transmitted through an antenna after amplification by the power amplifier (PA). In the receiver subsystem, the signal is passed through an on-chip low noise amplifier (LNA) and forwarded to non-coherent envelope detector. An adaptive comparator and baseband stages eliminate the duty cycle distortion and provide an error-free payload differential data stream at the output. The device maximises flexibility by offering OFF, IDLE, RECEIVER, and TRANSMITTER modes. The IDLE RX-sensing mode is optimised for very low power consumption and allows for automatic wake-up. In this mode, the receiver monitors and detects if any OOK modulation is present in order to wake-up the chip. Transition between different modes is achieved through direct configuration pins or through the $I^2C$ interface.

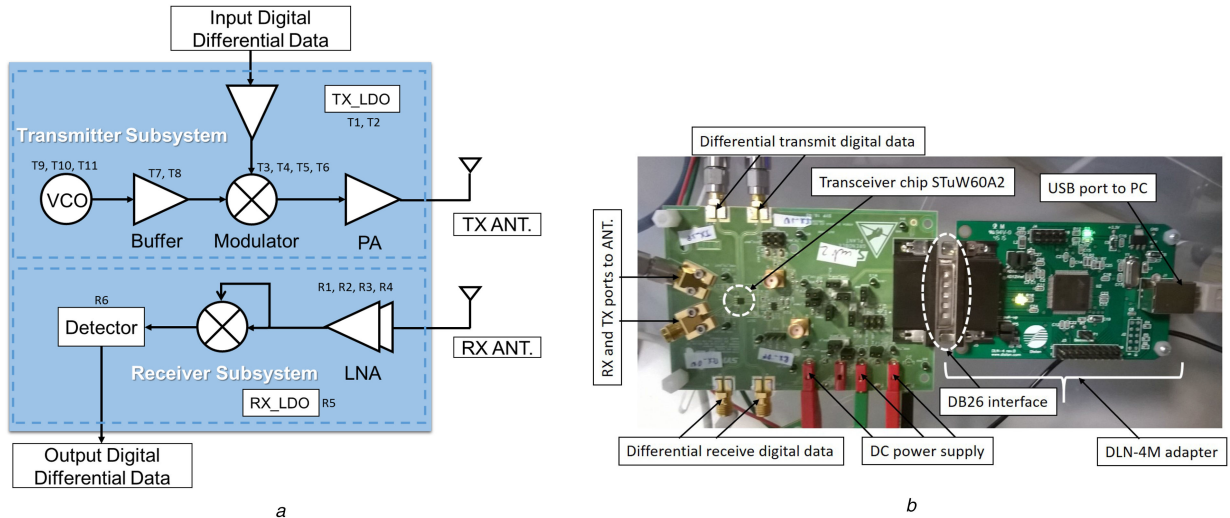

**Fig. 1** *60 GHz transceiver*
*(a)* Chip block diagram, where T1 - T11 and R1 - R6 show analogue voltage test points for transmitter and receiver subsystems, respectively, *(b)* Chip mounted on evaluation board, while DLN-4M adapter providing DB26 to USB interface

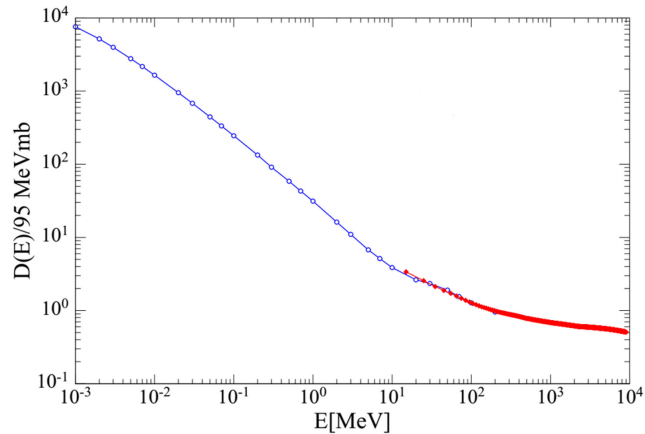

**Fig. 2** *Proton induced displacement damages in silicon, presented in [12, 13], are combined and distinguished with different colours*

### 2.1 Transceiver evaluation board

Two test PCBs with TRX chips and required drive electronics have been prepared for chip characterisation. As shown in Fig. 1*b*, the PCB has been provided with transmitter and receiver ports, as well as differential digital data ports. To monitor the power consumption of the chip separately, independent DC power ports have been made available for the chip and other drive electronics on the board. A DLN-4M adapter [https://diolan.com/dln4m] has been used for $I^2C$–USB interface. This allows the control and characterisation of the chip through an embedded software via PC.

## 3 Proton irradiation

The radiation damages in electronics for HEP experiment can be studied and approximated well with proton irradiation; even though protons will not be the most common particles in the HEP environment. The effects of proton irradiation can be scaled to and compared with the expected radiation environment in HEP experiments [7].

### 3.1 Irradiation damages

During irradiation, energy is deposited in semiconductors through electronic ionisation and atomic collisions, resulting in ionisation and displacement damages [8]. High ionising radiation doses damage oxides and oxide/semiconductor interfaces [9, 10], resulting a shift in threshold voltage, increased leakage current, and reduced charge mobility [11]. In addition, the high-energy particles collide with the semiconductor lattice atoms and create vacancies and interstitial sites in the lattice, resulting in defects called the displacement damages. For silicon, the relation between proton energy and displacement damage (D(E)), normalised to 1 MeV neutrons (95 MeVmb) is presented in Fig. 2. The curve shows that the protons being used in our study (17 MeV) are ~3.6 times more damaging than 1 MeV neutrons.

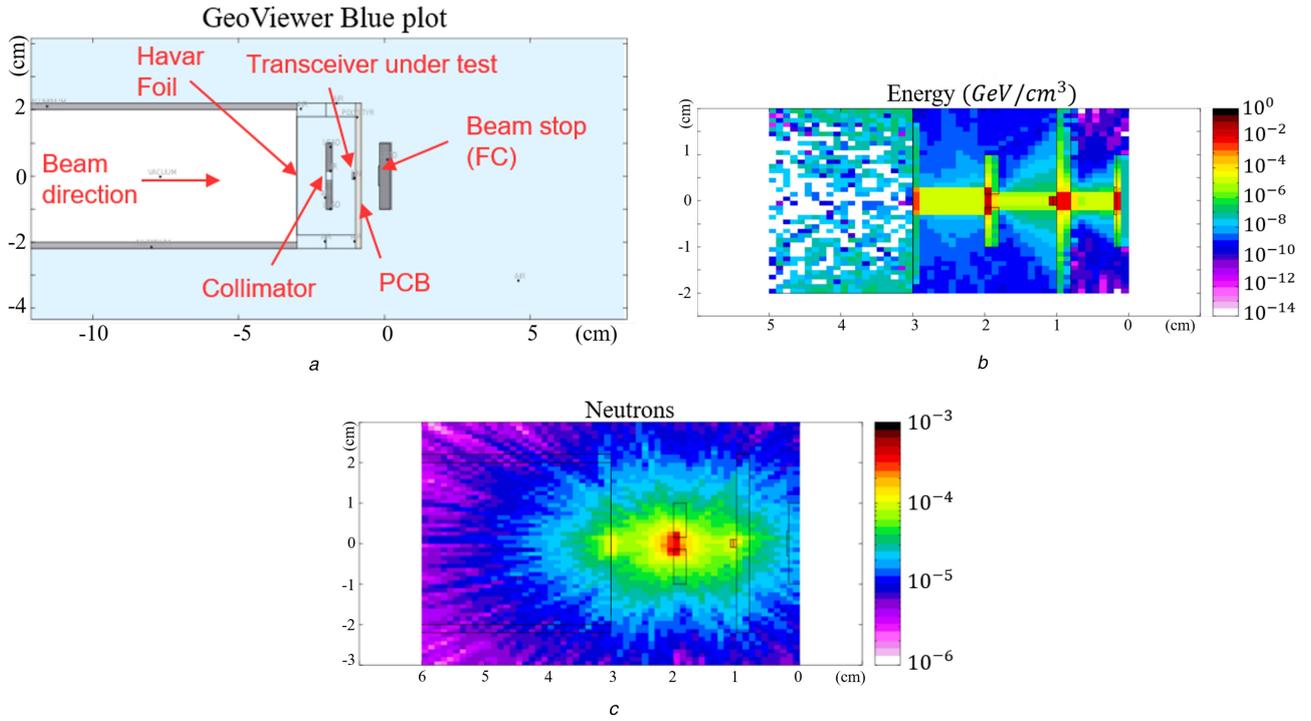

**Fig. 3** *Irradiation simulations using FLUKA*
*(a)* Simulation setup,
*(b)* Total ionising dose (TID) per incoming proton,
*(c)* Neutron yield per incoming proton

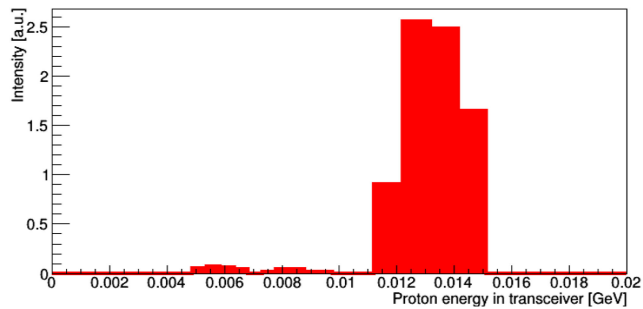

**Fig. 4** *Simulated proton spectrum in irradiated silicon transceiver chip*

## 3.2 Simulations and calculations

The software package FLUKA [http://www.fluka.org] has been used for simulations of particle transport and interactions with matter. As shown in the simulation setup in Fig. 3*a*, a proton beam is extracted from the cyclotron through a Havarfoil [http://www.goodfellow.com/E/Havar-High-Strength-Non-Magnetic-Alloy-Foil.html] window. A collimator is used to restrict the beam to an area of $3 \times 3$ mm$^2$ to protect the surrounding electronics on the PCB. Both collimator and beam stop are made of lead (Pb). Given the unknown radiation hardness of the TRX chip, an initial fluence ($F$) of about $10^{14} N_{eq}/$ cm$^2$ was selected. Fig. 3*b* shows the TID per incident proton while neutron distribution is shown in Fig. 3*c*. These neutrons are secondary particles that could potentially damage electronics surrounding the TRX chip. It is further shown in Fig. 4 that a significant amount of energy is lost by the protons after passing through the TRX chip, indicating considerable energy transfer to the DUT.

The simulated energy deposited in the TRX chip, $E_{sim}$, is 28 MeV/cm$^3$ per incoming proton, resulting in 192 kGy TID. To calculate the required time to achieve the target fluence ($F$) of $10^{14} N_{eq}/$ cm$^2$, we assume 1 nA beam current, which gives $6.25 \times 10^9$ protons/s. As the collimator has restricted the irradiation area to $3 \times 3$ mm$^2$ (~0.1 cm$^2$), with planned beam current, the required fluence from the accelerator can be achieved in

$$T_{irad} = F \times A_{irad}/(S \cdot F \cdot \times N_p) = 444 \text{ sec } (7.4 \text{ min}) \quad (1)$$

where $F$ = Target fluence $(10^{14} N_{eq}/$ cm$^2)$, $A_{irad}$ = Effective irradiation area (0.1 cm$^2$), S · F · = Damage Scaling Factor (3.6), and $N_p$ = No. of protons per second for 1 *nA* beam current ($6.25 \times 10^9$).

## 4 Irradiation experiment

The irradiation experiment was performed at Åbo Akademi University, Turku, Finland. Fig. 5*a* shows the block diagram of the measurement setup in the lab. A Stratix V FPGA board was used to generate a pseudo-random binary sequence 7 (PRBS7) at 5 Gbps. This data was sent as a differential pair to the transmitter TRX (TX), which further transmitted the 60 GHz modulated signal towards the receiver TRX (RX) module. To simplify the experimental setup, a coaxial cable connection was used between output of the TX board and input of the RX board, as opposed to an over-the-air link with antennas. The RX forwarded the detected signal in differential pair to the Stratix V FPGA for data testing. The cyclotron (MGC-20) was set to 17 MeV protons and the TRX evaluation board operating in RX mode was placed on the beam line attached to the irradiation chamber with an air propagation distance of 30 cm between the beam aperture and the chip. As shown in Fig. 5*b*, the board is fitted between the collimator plates

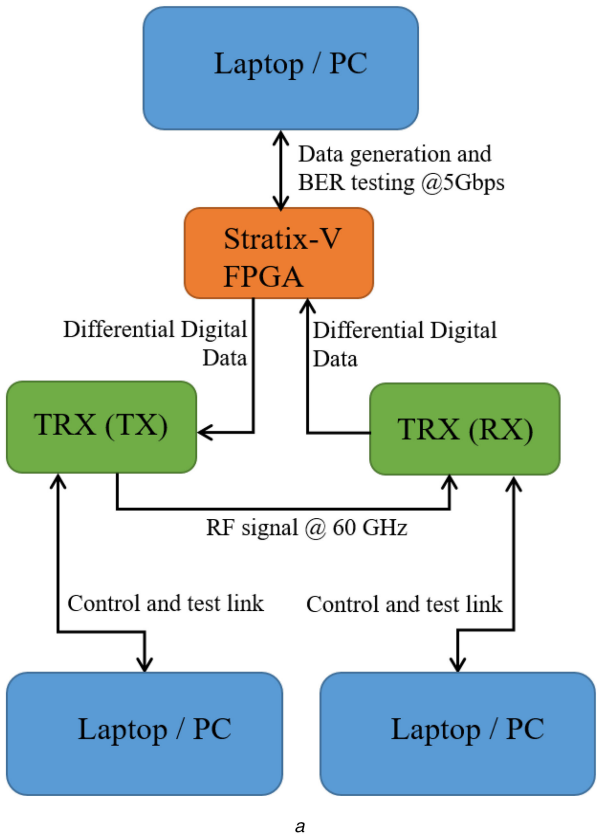
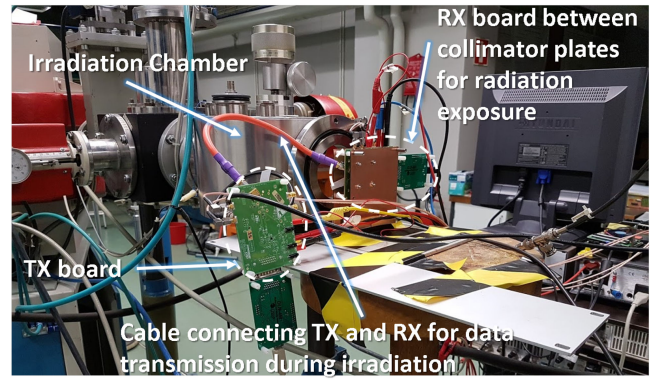

**Fig. 5** *Measurement setup at cyclotron facility*
*(a)* Block diagram of the setup at cyclotron facility, RX and TX were in turn exposed to high energy beam,
*(b)* Experimental setup at Åbo Akademi University, Finland

**Table 1** Calculated fluences and doses during irradiation

| DUT | Calculated fluence ($N_{eq}/cm^2$) | Calculated TID (*kGy*) |
| --- | --- | --- |
| RX | $1.38 \times 10^{14}$ | 74 |
| TX | $0.78 \times 10^{14}$ | 42 |

and both boards are connected by the coaxial cable, during the irradiation.

Additionally, the film dosimeter was placed on top of the TRX to record the cumulative dose. The RX chip was irradiated first, as it requires a pristine TX chip to test, while the chip can be evaluated in TX mode in isolation, without the RX. Each irradiation experiment was done in three steps with increasing beam current. Before the irradiation, a calibration experiment was performed with only the irradiation setup (collimator and beam stop) to determine the beam current. These beam current calibrations were used for the fluence and dose calculations given in Table 1. The TID and fluence values are lower for the TX because the TRXs were characterised multiple times during the irradiation, which made the cyclotron filament move out of apex towards the end of the experiment when the TX was being exposed.

### 4.1 Pre- and post-irradiation results and discussion

After the experiment, both TRX PCBs were stored in a freezer for 3 weeks to minimise the annealing, while waiting for the activity to decay. In the post-irradiation analysis, both the RX and the TX have been found functional through over-the-air measurements at 5 Gbps. The comparison of pre- and post-irradiation analogue voltage measurements for the test points shown in Fig. 1*a*, is tabulated in Table 2. For the TX, low dropout voltages (T1 & T2), mixer voltages (T3 & T4), modulator voltages (T5 & T6), buffer voltages (T7 & T8), VCO voltages (T9, T10 & T11) and bandgap voltage (VBG) is measured. While for the RX, two stage LNA voltages (R1, R2, R3 & R4), baseband LDO voltage (R5), gate bias detector voltage (R6), amplitude control signal voltage (VATC) and bandgap voltage (VBG) are tabulated. It can be observed that there is no significant difference in the pre- and post-irradiation measurements for the TX except the VCO voltage T9, but the RX chip looks comparatively more affected by the irradiation. This is in accordance with the provided TID and fluence. The measurements show that the RX VBG circuit, which provides voltage references to different blocks of the chip, remained unaffected by the irradiation. At output of the VBG, a PTAT (proportional to absolute temperature) circuit is used that generates a reference current for the RF blocks to compensate the temperature variations. The implemented PTAT circuit is based on two bipolar transistors, whose junctions are supposedly affected by the ionisation dose. This possible degradation of the current reference circuit is seen as voltage variations in the LNA stages of the RX chip.

Likewise, a power consumption comparison is presented in Table 3. The measurements were taken for an active transmission link (Data-ON) and without data transmission (Data-OFF). These results also show notable differences in pre- and post-irradiation measurements for the RX compared to the TX. The reduction in the power consumption can be caused by radiation induced traps in the $SiO_2/Si$ interfaces. It has been reported that the absolute value of the threshold voltage increases in both *p*- and *n*-channel MOSFETs due to creation of new interface traps [9]. This increase in the threshold voltage is a possible reason of reduced power consumption, which in turn resulted a reduction in the TX output power and the RX gain by 1 and 4.5 dB, respectively. Sufficient link budgeting can fully compensate the reduction in the TX power, while up to 2 dB loss in the RX gain can also be recovered.

**Table 2** Pre- and post-irradiation analogue test comparison at 5 Gbps

| Transmitter (TX) | | | Receiver (RX) | | |
|---|---|---|---|---|---|
| Test | Pre (V) | Post (V) | Test | Pre (V) | Post (V) |
| LDO1 (T1) | 1.006 | 1.007 | VDDA (R1) | 1.116 | 1.079 |
| LDO2 (T2) | 1.207 | 1.208 | VGGA (R2) | 0.558 | 0.497 |
| VDDA (T3) | 1.007 | 1.008 | VDDA (R3) | 1.115 | 1.077 |
| VGGA (T4) | 0.710 | 0.706 | VGGA (R4) | 0.555 | 0.491 |
| VDDA (T5) | 0.990 | 0.992 | LDO-BB (R5) | 1.219 | 1.177 |
| VGGA (T6) | 0.517 | 0.517 | VGDET (R6) | 0.215 | 0.216 |
| VDDA (T7) | 0.993 | 0.995 | VATC | 0.972 | 0.940 |
| VGGA (T8) | 0.509 | 0.509 | VBG (PM) | 1.208 | 1.208 |
| VDDA (T9) | 0.516 | 0.563 | — | — | — |
| VGGA (T10) | 0.633 | 0.636 | — | — | — |
| Vtune (T11) | 0.669 | 0.661 | — | — | — |
| VBG (PM) | 1.206 | 1.207 | — | — | — |

**Table 3** Pre- and post-irradiation power consumption for TX and RX

| Test Name | Pre-irradiation | Post-irradiation |
|---|---|---|
| TX-Data-OFF | 33.9 mW | 32.4 mW |
| TX-Data-ON | 40.6 mW | 43 mW |
| RX-Data-OFF | 15.2 mW | 11.4 mW |
| RX-Data-ON | 14.7 mW | 7.8 mW |

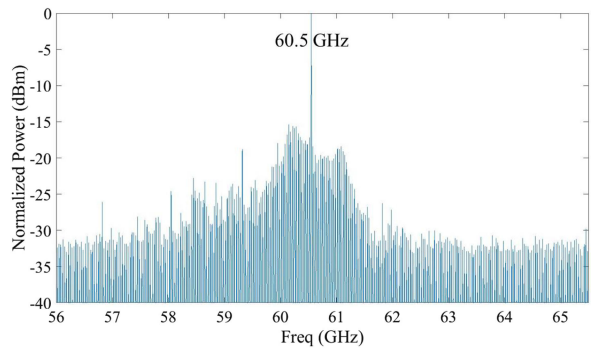

**Fig. 6** *Normalised power spectrum of 5 Gbps data transmission while transmitter was set on continuous wave mode. Highest peak is at 60.5 GHz*

In addition, the RX envelope detector and digital interface remained unaffected by the irradiation.

A spectrum analyser [Keysight Technologies 11970V 50-75 GHz Series Harmonic Mixer] with external V-band mixer was used in the pre-, post- and during-irradiation characterisations to monitor the RF power level as a function of frequency. The normalised power spectrum of 5 Gbps data transmission is shown in Fig. 6. The transmitter was set on continuous wave mode and a small downshift of 10 MHz in the centre frequency was observed after the irradiation. This is in agreement with the results from RD-53 collaboration [14] at CERN. From their studies, we can expect higher doses to shift the central frequency even more. However, as the chip uses the envelope detection in RX mode, this shift will not affect the communication.

The test setup for wireless data measurement at 5 Gbps with the help of horn antennas is shown in Fig. 7*a*. The corresponding eye diagram with the RMS jitter of 21 ps, 60 mV eye height and 67 ps eye width is presented in Fig. 7*b*. A similar level of jitter was obtained in post-irradiation measurements with comparatively reduced range. Similar findings are presented in Table 4 as pre- and post-irradiation bit error rate (BER) test results at 1.25 Gbps for $10^{12}$ tested bits. The higher post-irradiation BER can be a result of increased white noise and $1/f$ noise. The reduced charge mobility leads to degradation in transconductance ($g_m$), thus giving rise to the white noise, while capture and emission of charges from radiation induced border traps gives rise to frequency-dependent $1/f$ noise [15]. Moreover, 5.5 dB degradation in the link budget is also responsible for higher BER.

## 5 Conclusion and future work

The radiation hardness of a V-band TRX chip in 65 nm CMOS, acting both in TX and RX modes, was investigated after 17 MeV proton irradiation. TID and fluence used during the experiment were 74 & 42 kGy and $1.4 \times 10^{14} N_{eq}/$ cm$^2$ & $0.8 \times 10^{14} N_{eq}/$ cm$^2$ for the RX & the TX, respectively. The RX experienced higher loss of sensitivity than the TX power loss, which is in accordance with the respective TID and fluence. However, both the RX and the TX were found operational through over-the-air measurements at 5 Gbps after irradiation with losses that can partially be compensated through link budgeting.

Although the radiation levels expected in the real experiment at LHC are much higher, this experiment presents an initial investigation of radiation hardness that gives encouraging results despite the fact that the devices were not specifically radiation hardened by either design or process. In future, the TRX ICs will be tested at higher doses to analyse the technological limitations.

## 6 Acknowledgments

The authors would like to thank Pierre Busson, Nicolas Ricome and Frederic Lagrange from STMicroelectronics for providing equipment and support as well as Åbo Akademi University, Finland for use of their irradiation facility. Imran Aziz is thankful to MUST, Pakistan for their financial assistance and Sebastian Dittmeier acknowledges support by the International Max Planck Research School for Precision Tests of Fundamental Symmetries.

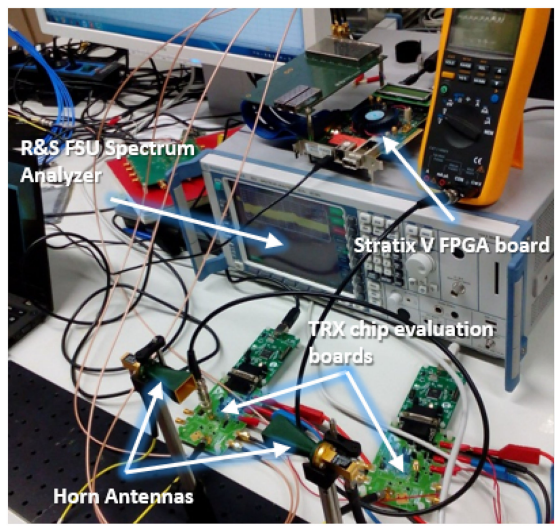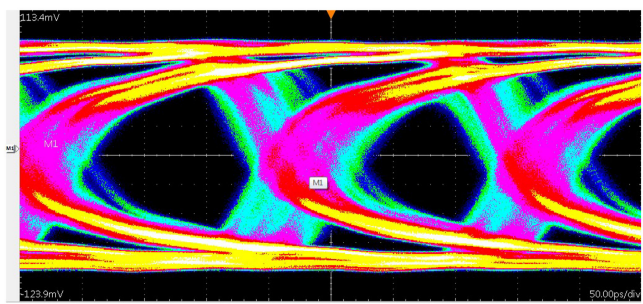

*a*

*b*

**Fig. 7** *5 Gbps wireless measurements*
*(a)* Point-to-point wireless link established with horn antennas,
*(b)* Eye diagram from Tektronix DSA8300 Digital Serial Analyser

**Table 4** Pre- and post-irradiation wireless BER test for $10^{12}$ tested bits

| Distance, cm | Pre-irradiation BER | Post-irradiation BER |
| --- | --- | --- |
| 6 | 0 | 0 |
| 8 | 0 | $1.4 \times 10^{-10}$ |
| 10 | 0 | $9.0 \times 10^{-6}$ |